\documentclass[prd,twocolumn,nofootinbib,nobibnotes,noeprint,preprintnumbers,floatfix]{revtex4-2}
\usepackage{amssymb,amsmath,graphicx}
\usepackage{subcaption}
\usepackage{bm}
\usepackage[dvipsnames]{xcolor}
\usepackage[colorlinks,allcolors=NavyBlue]{hyperref}
\usepackage{microtype}

\begin{document}

\preprint{FERMILAB-PUB-23-272-T}

\title{Diffuse Ultra-High-Energy Gamma-Ray Emission From TeV Halos}

\author{Ariane Dekker$^2$}
\thanks{ahdekker@uchicago.edu, http://orcid.org/0000-0002-3831-9442}

\author{Ian Holst$^{1,2}$}
\thanks{holst@uchicago.edu, https://orcid.org/0000-0003-4256-3680}

\author{Dan Hooper$^{1,2,3}$}
\thanks{dhooper@fnal.gov, http://orcid.org/0000-0001-8837-4127}

\author{Giovani Leone$^{2,4}$}
\thanks{grleone@uchicago.edu, http://orcid.org/0000-0001-7328-7397}

\author{Emily Simon$^1$}
\thanks{ersimon@uchicago.edu, http://orcid.org/0000-0002-9386-630X}

\author{Huangyu Xiao$^3$}
\thanks{huangyu@fnal.gov, http://orchid.org/0000-0003-2485-5700}

\affiliation{$^1$University of Chicago, Department of Astronomy and Astrophysics, Chicago, Illinois 60637, USA}
\affiliation{$^2$University of Chicago, Kavli Institute for Cosmological Physics, Chicago, Illinois 60637, USA}
\affiliation{$^3$Fermi National Accelerator Laboratory, Theoretical Astrophysics Group, Batavia, Illinois 60510, USA}
\affiliation{$^4$University of Chicago, Department of Physics, Chicago, Illinois 60637, USA}

\date{\today}

\begin{abstract}
The LHAASO Collaboration has recently reported a measurement of the diffuse gamma-ray emission from the Galactic Plane at energies between 10 TeV and 1 PeV. While this emission is brighter than that expected from cosmic-ray interactions in the interstellar medium alone, we show that the intensity, spectrum, and morphology of this excess are in good agreement with that predicted from the ``TeV halos'' which surround the Milky Way's pulsar population. These results support the conclusion that TeV halos dominate the ultra-high-energy sky, and that these objects convert $\sim 5\%$ of their total spindown power into very-high and ultra-high-energy photons.
\end{abstract}

\maketitle

\section{Introduction}

The Large High Altitude Air Shower Observatory (LHAASO) Collaboration has recently reported their measurement of the diffuse gamma-ray emission from the Galactic Plane at energies between 10 TeV and 1 PeV~\cite{LHAASO:2023gne}\footnote{Note that the LHAASO measurements are potentially in tension with earlier Tibet AS$\gamma$ results \cite{TibetASgamma:2021tpz}.}. Interestingly, the flux measured by LHAASO's square kilometer array (LHAASO-KM2A) exceeds that predicted from the interactions of cosmic rays in the interstellar medium (ISM) by a factor of $\sim 2-3$. Although this result could be the consequence of spatial variations in the distribution of cosmic rays~\cite{Guo:2018wyf,Lipari:2018gzn} or in the dust-to-gas ratio~\cite{Giannetti_2017}, it is more likely an indication that the diffuse ultra-high-energy gamma-ray flux receives significant contributions from unresolved sources~\cite{zhang2023galactic}. In particular, we will argue that ``TeV halos'' contribute significantly to this flux and are responsible for much of the Milky Way's diffuse ultra-high-energy gamma-ray emission.

In 2017, the HAWC Collaboration reported the observation of very-high-energy gamma-ray emission from the regions surrounding the Geminga and Monogem pulsars~\cite{HAWC:2017kbo} (see also Ref.~\cite{HAWC:2020hrt,Abeysekara:2017hyn}). The intensity and spectrum of this emission imply that these sources convert a significant fraction (on the order of 10\%) of their total spindown power into very high-energy electron-positron pairs, which go on to produce gamma rays through inverse Compton scattering. Furthermore, the spatial extent of these TeV halos (at a level of $\sim 2^{\circ}$, corresponding to $\sim 20 \, {\rm pc}$), indicates that cosmic rays propagate much less efficiently in the vicinity of these pulsars than they do elsewhere in the ISM~\cite{Hooper:2017gtd,Hooper:2017tkg,Johannesson:2019jlk,DiMauro:2019hwn,Liu:2019zyj} (for earlier work see \cite{2000A&A...362..937A}).

Since the discovery of Geminga and Monogem's extended multi-TeV emission, it has become clear that most, and perhaps all, middle-aged ($t\sim 10^5-10^6 \, {\rm yr}$) pulsars are surrounded by a TeV halo~\cite{Linden:2017vvb,Sudoh:2019lav,Sudoh:2021avj}.\footnote{There is also evidence for TeV halos associated with millisecond pulsars~\cite{Hooper:2021kyp,Hooper:2018fih,LHAASO:2023rpg}.} In particular, a large fraction of the sources detected by HAWC~\cite{HAWC:2020hrt,Abeysekara:2017hyn,HAWC:2021dtl,HAWC:2019tcx} and LHAASO~\cite{LHAASO:2023rpg} (and many detected by HESS~\cite{HESS:2018pbp,HESS:2017lee}) are spatially coincident with a pulsar. Moreover, there exists a strong correlation between a pulsar's spindown power and its observed ultra-high-energy gamma-ray luminosity.

In this paper, we argue that the excess of ultra-high-energy diffuse gamma-ray emission observed from the Galactic Plane is likely generated by a large population of unresolved TeV halos. We employ a Monte Carlo simulation to model the Milky Way's pulsar population and predict the ultra-high-energy gamma-ray emission from the associated TeV halos. We compare this to the measurements reported by the LHAASO Collaboration, finding that the spectrum, morphology, and overall intensity of this signal are in good agreement with that expected from TeV halos.

\section{Modeling the Milky Way's TeV Halo Population}

\begin{figure*}
    \centering
    \includegraphics[width=\textwidth]{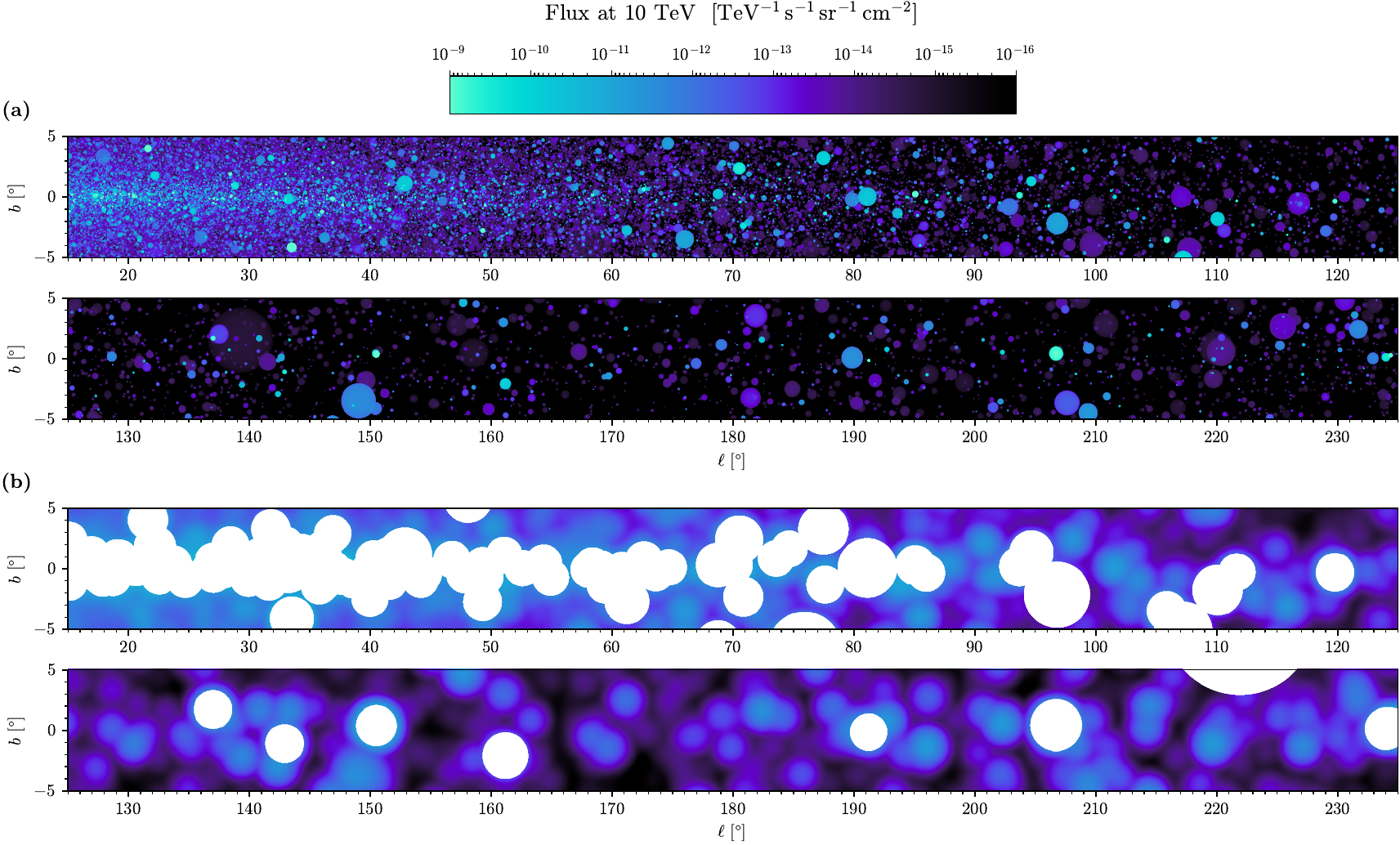}
    \caption{A realization of our Monte Carlo simulation of the ultra-high-energy gamma-ray emission from TeV halos, before (a) and after (b) convolving the map with LHAASO-KM2A's point spread function and applying a mask around the brightest sources. Both (a) and (b) are subdivided into the inner galaxy (top) and outer galaxy (bottom) regions.}
    \label{fig:skymap}
\end{figure*}

To study the possibility that the diffuse ultra-high-energy gamma-ray emission from the Galactic Plane receives a significant contribution from TeV halos, we have developed a Monte Carlo simulation to model the characteristics of this source population. In our calculations, we take the gamma-ray luminosity of a given TeV halo to be proportional to its spindown power, $L_{\gamma} = \eta \, \dot{E}_{\rm rot}$, where $\eta$ is the gamma-ray efficiency. The spindown power of a pulsar can be expressed as follows:
\begin{align}
\dot{E}_{\rm rot} &= -\frac{4\pi^2 I \dot{P}}{P^3} \\
&= -\frac{4 \pi^2 I}{P_0^2 (n-1) \tau} \bigg(\frac{t}{\tau}+1\bigg)^{-(n+1)/(n-1)}, \nonumber
\end{align}
where $I$ is the neutron star's moment of inertia, $P$ is its period, $P_0$ is its period at birth, and $\tau = 3 c^3 I P_0^2/4\pi^2 B^2 r^6$ is its spindown timescale, where $B$ and $r$ are the neutron star's magnetic field and radius. The quantity $n$ is the pulsar's braking index, defined as $\dot{P} \propto P^{-n+2}$. Note that $n=3$ corresponds to the case in which pulsars lose energy through magnetic dipole braking.

In our Monte Carlo simulation, we select a random age for each pulsar (between $t=2\times 10^4$ and $10^8 \, {\rm yr}$) and use this to determine the pulsar's spindown power, adopting parameters of $n=3$ and $\tau =10^4 \, {\rm yr}$, and normalizing this relationship such that $\dot{E}_{\rm rot} \approx -1.1 \times 10^{34} \, {\rm erg/s} \times  (t/10^{6} \, {\rm yr})^{-2}$ (corresponding to the default parameters adopted in Ref.~\cite{Bitter:2022uqj}). We take the Milky Way's pulsar birth rate to be 0.7 per century. For the location of each pulsar, we draw from the
following spatial distribution~\cite{Lorimer:2003qc}:
\begin{align}
n_{\rm pulsar} \propto  R^{2.35} \,  e^{-R/1530 \, {\rm pc}} \, e^{-|z|/z_s},
\end{align}
where $R$ and $z$ describe the location of a given pulsar in cylindrical coordinates, and the Solar System is located at $R= 8.12 \, {\rm kpc}$~\cite{GRAVITY:2018ofz} and $z = 0.021 \, {\rm kpc}$~\cite{Bennett_2018} (we take this distribution to be symmetric with respect to the angular coordinate). For the vertical scale height, we adopt $z_s = 70 \, {\rm pc} + 180 \, {\rm pc} \times (t/10^6 \, {\rm yr})$, up to a maximum value of $z_s=1 \, {\rm kpc}$.  The $z_s = 70 \, {\rm pc}$ term in this expression accounts for the distribution of the progenitors~\cite{2004A&A...425.1009M}, while the age-dependent term reflects the broadening that results from natal kicks~\cite{Hansen:1997zw}.

From our simulated pulsar population, we calculate a sky map of the resulting gamma-ray emission as a function of energy. We adopt a power-law spectrum $dN_{\gamma}/dE_{\gamma} \propto E^{-\alpha}_{\gamma}$ for each source, finding that an index of $\alpha = 3.1$ between energies of $E_\mathrm{min} = 1\,\mathrm{TeV}$ and $E_\mathrm{max} = 1\,\mathrm{PeV}$ fits the data well and is consistent with the highest-energy observations of TeV halos~\cite{HAWC:2019tcx}. In generating our simulated sky map, we take each TeV halo to have a radius $R_\mathrm{halo} = 10 \, {\rm pc}$ \cite{HAWC:2017kbo}, and to produce emission uniformly within the corresponding sphere. The photon flux per solid angle for each halo is given by
\begin{equation} \label{eq:profile}
    I(E_\gamma,\theta) = \frac{3 (\alpha-2) \eta \dot{E}_\mathrm{rot}}{8 \pi^2}\, \frac{E_\gamma^{-\alpha}}{E_\mathrm{min}^{-\alpha+2}}\, \frac{\sqrt{R_\mathrm{halo}^2 - (\theta d)^2}}{R_\mathrm{halo}^3},
\end{equation}
within an angle $\theta < R_\mathrm{halo} / d$ of the pulsar's location, where $d$ is the distance to the pulsar. The details of the angular emission profile, including the size, do not significantly affect our results.

Gamma rays can be attenuated via pair production in scattering with Galactic photon fields. To calculate the optical depth of the Milky Way, we model the radiation fields of the ISM as a sum of blackbodies associated with the cosmic microwave background ($\rho_{\rm CMB} = 0.260 \, {\rm eV/cm}^3$, $T_{\rm CMB}=2.7 \, {\rm K}$), starlight ($\rho_{\rm star} = 0.60 \, {\rm eV/cm}^3$, $T_{\rm star}=5000 \, {\rm K}$), and infrared emission ($\rho_{\rm IR} = 0.60 \, {\rm eV/cm}^3$, $T_{\rm IR}=20 \, {\rm K}$)~\cite{Porter:2017vaa}. Although gamma-ray attenuation is negligible at energies below $\sim 100 \, {\rm TeV}$, we find that pair production reduces the total flux predicted from TeV halos by \mbox{$\sim 30-40\%$} at $E_{\gamma} = 1 \, {\rm PeV}$.

We convolve our sky map with a $\sigma_{\rm PSF}=0.6^{\circ}$ Gaussian to account for LHAASO-KM2A's point spread function, and apply a mask to the brightest of our simulated sources, based on the characteristics of the mask adopted by the LHAASO Collaboration~\cite{LHAASO:2023gne}. More specifically, we mask all sources in the convolved map whose total flux at 10 TeV exceeds $5 \times 10^{-15} \, {\rm TeV}^{-1} \,  {\rm cm}^{-2} \,   {\rm s}^{-1}$, removing circular regions around each such source out to a radius of $2.5 \times (\sigma_{\mathrm{halo}}^2+\sigma_{\rm PSF}^2)^{1/2}$, where $\sigma_{\mathrm{halo}}$ is the angular radius of the TeV halo.

\section{Results}

\begin{figure}
    \centering
    \includegraphics[width=\columnwidth]{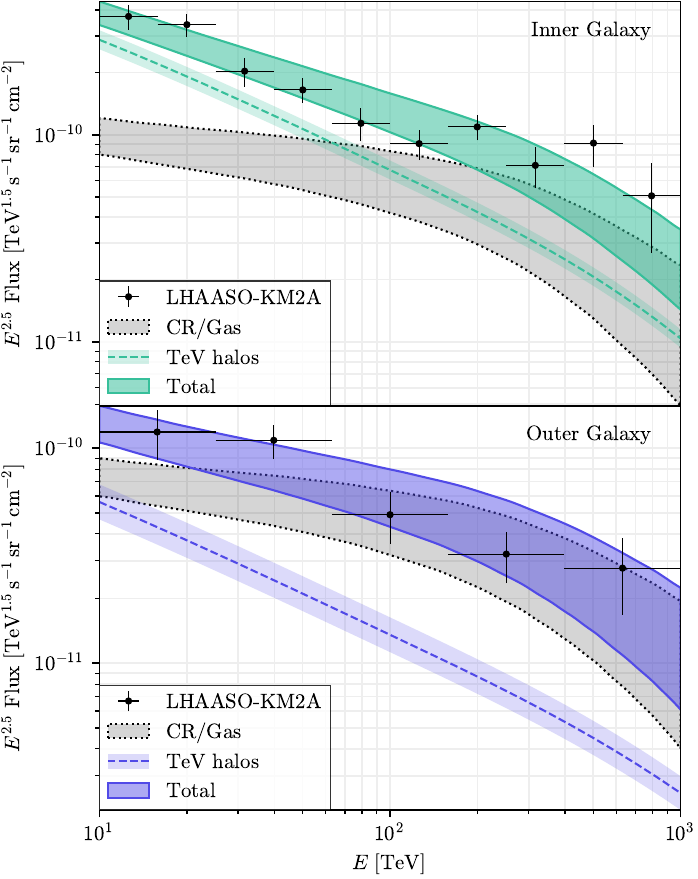}
    \caption{The spectrum of the diffuse ultra-high-energy gamma-ray emission averaged over the unmasked portions of the inner ($15^{\circ}<l < 125^{\circ}$, $-5^{\circ}<b<5^{\circ}$) and outer ($125^{\circ}<l < 235^{\circ}$, $-5^{\circ}<b<5^{\circ}$) Galaxy. The spectrum reported by the LHAASO Collaboration~\cite{LHAASO:2023gne} is compared to that predicted from cosmic-ray interactions with gas in the interstellar medium, from TeV halos, and from the sum of these contributions. The combined emission from TeV halos and cosmic-ray interactions is in good agreement with the observed spectrum.}
    \label{fig:spectrum}
\end{figure}

In Fig.~\ref{fig:skymap}, we show one realization of our simulated sky map, over the same regions of the Galactic Plane considered by the LHAASO Collaboration~\cite{LHAASO:2023gne}, and evaluated at an energy of $E_{\gamma}=10 \, {\rm TeV}$. To normalize our results, we have adopted an efficiency of $\eta=0.052$, defined as the gamma-ray luminosity (integrated above 1 TeV) divided by the total spindown power. Note that the value of $\eta$ is degenerate with $E_\mathrm{min}$ and with some of the pulsar spindown parameters. The masking threshold also affects the normalization of the results and thus is also degenerate with $\eta$.

In Fig.~\ref{fig:spectrum}, we show the spectrum of the simulated emission from TeV halos averaged over the unmasked portions of the inner ($15^{\circ}<l < 125^{\circ}$, $-5^{\circ}<b<5^{\circ}$) and outer ($125^{\circ}<l < 235^{\circ}$, $-5^{\circ}<b<5^{\circ}$) Galaxy as colored dashed lines. The width of the bands around them reflects the variations obtained over 10 different realization of our Monte Carlo. This result is compared to the spectrum reported by the LHAASO Collaboration~\cite{LHAASO:2023gne} (black points), and to that predicted from cosmic-ray interactions with gas in the ISM. The cosmic ray predictions (grey band) are also taken from Ref~\cite{LHAASO:2023gne}, which assumes an uniform cosmic-ray spectrum and gas column densities as inferred from the PLANCK dust opacity map~\cite{Planck:2016frx} with a constant dust-to-gas ratio for the spatial template, consistent with the assumptions reported by the LHAASO collaboration. Moreover, the width of the grey band represents the uncertainties in the cosmic ray spectra, which is obtained by proton and helium measurements based on balloon, space and ground-based air shower experiments~\cite{LHAASO:2023gne}. Notably, the spectrum measured by LHAASO-KM2A exceeds that predicted from cosmic-ray interactions by a factor of $\sim 3$ in the inner Galaxy region, and by a factor of $\sim 2$ in the outer Galaxy region (at least at energies below $\sim 60 \, {\rm TeV}$). The upper solid colored band in each frame represents the sum of the contributions from TeV halos and cosmic ray interactions, which is in good agreement with the spectrum reported by the LHAASO Collaboration.

\begin{figure*}
    \centering
    \includegraphics[width=\textwidth]{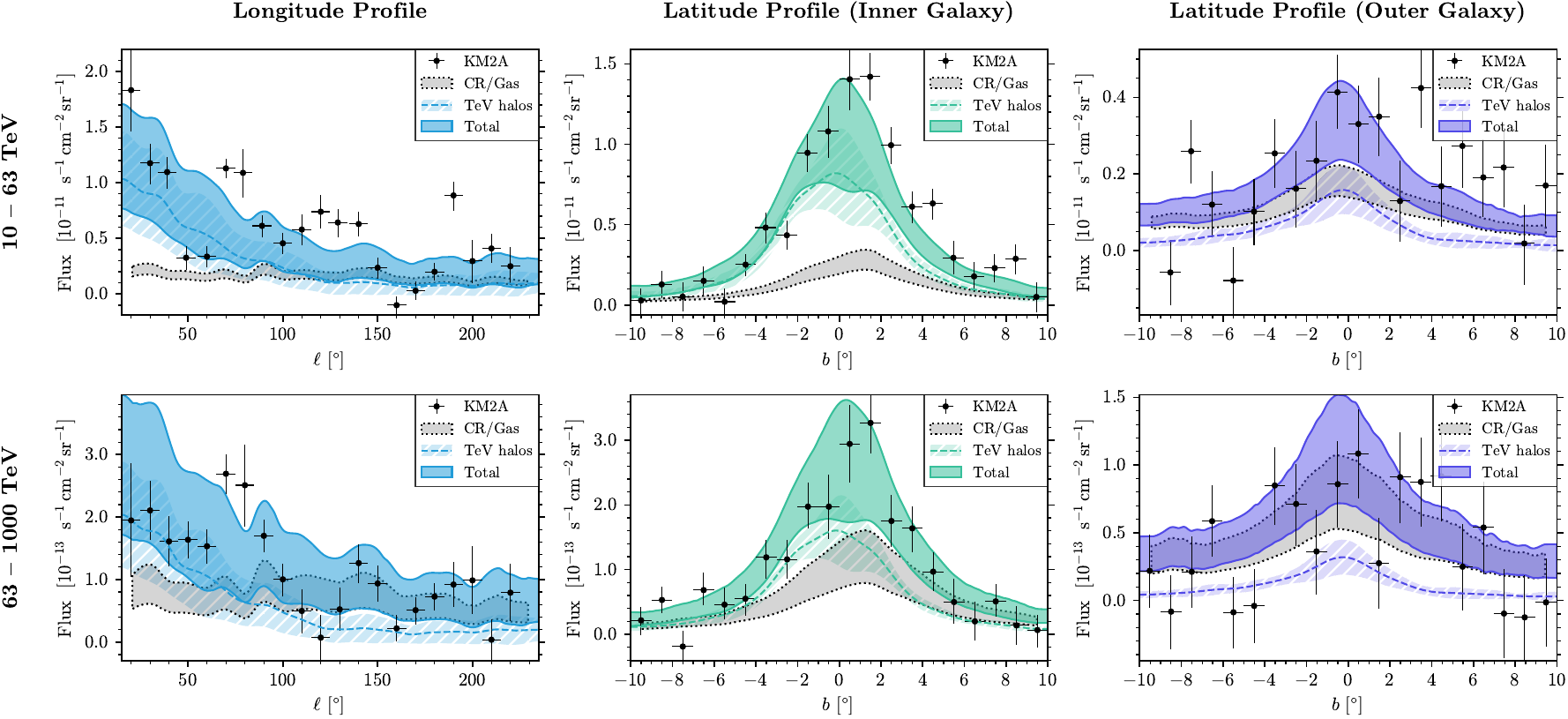}
    \caption{Longitude and latitude profiles of the diffuse ultra-high-energy gamma-ray emission measured by the LHAASO Collaboration~\cite{LHAASO:2023gne}, compared to the predictions from cosmic-ray interactions with gas, from TeV halos, and from the sum of these contributions. In the left, center, and right frames, the flux has been integrated over $-5^{\circ}<b<5^{\circ}$, $15^{\circ}<l<125^{\circ}$, and $125^{\circ}<l<235^{\circ}$, respectively. The top (bottom) frames show the fluxes as integrated between $10-63 \, {\rm TeV}$ ($63-1000 \, {\rm TeV}$). The combined emission from TeV halos and cosmic-ray interactions is in good agreement with the observed spectrum.}
    \label{fig:profiles}
\end{figure*}

In Fig.~\ref{fig:profiles}, we show the longitude and latitude profiles of the diffuse emission measured by LHAASO-KM2A, once again comparing this to that predicted from cosmic-ray interactions with gas and from TeV halos. In generating these TeV halo bands, we have convolved the profiles with a $0.5^\circ$ or $5^\circ$ Gaussian, reflecting the bin width of the LHAASO data points~\cite{LHAASO:2023gne}. Note that, unlike in Ref.~\cite{LHAASO:2023gne}, we have {\it not} allowed the magnitude of the cosmic-ray component to float in producing these figures. The curves are instead normalized to the same values that were used in Fig.~\ref{fig:spectrum}.

\section{Discussion and Summary}

The TeV halos that surround our Galaxy's pulsar population are expected to produce a significant fraction of our galaxy's very-high and ultra-high-energy gamma-ray emission. In this context, it was argued by Linden and Buckman~\cite{Linden:2017blp} that the diffuse TeV-scale emission observed from the Galactic Plane by Milagro~\cite{Milagro:2005xqq} could be attributed to TeV halos (see also~\cite{Vecchiotti_2022,Yan:2023uxd,Martin:2022aun}). Here, we have considered the diffuse gamma-ray emission from the Galactic Plane at energies between 10 TeV and 1 PeV, as reported by the LHAASO Collaboration~\cite{LHAASO:2023gne}. This emission cannot be easily accounted for with cosmic-ray interactions alone. Although previous studies have made a case based on spectral characteristics, we have now shown that the angular morphology is also in good agreement with that expected from TeV halos.

The intensity of the diffuse flux measured by LHAASO-KM2A indicates that approximately $\sim 5\%$ of the total spindown power of TeV halos goes into the production of very-high and ultra-high-energy gamma-ray emission. This efficiency is consistent with previous observations of individual TeV halos~\cite{Hooper:2017gtd,Linden:2017vvb,Sudoh:2019lav,Sudoh:2021avj}, and with that required for TeV halos to generate the observed cosmic-ray positron excess~\cite{Hooper:2017gtd,Fang:2018qco,Profumo:2018fmz,Tang:2018wyr,Manconi:2020ipm,Bitter:2022uqj}.

The results presented in this paper support the conclusion that TeV halos dominate the inner Milky Way's ultra-high-energy gamma-ray emission. Furthermore, much of this emission originates from individual TeV halos that are not far below the detection thresholds of existing gamma-ray telescopes. This strongly suggests that future very-high and ultra-high-energy gamma-ray telescopes will be capable of detecting large numbers of currently unresolved TeV halos~\cite{Sudoh:2019lav,Linden:2017vvb}.

\vspace{6pt}

Note: While this paper was in preparation, the LHAASO source catalog \cite{LHAASO:2023rpg} identified many sources of multi-TeV emission spatially correlated with known pulsars, as would be expected based on the results of our study. More recently, the IceCube collaboration reported the detection of neutrinos from the galactic plane, suggesting at least part of this diffuse emission has a hadronic origin \cite{IceCube:2023ame}. With the information currently available, it is difficult to make a robust determination of this fraction. We leave this question to future work.

\vspace{6pt}

\textbf{Acknowledgments.} AD is supported by the Kavli Institute for Cosmological Physics at the University of Chicago through an endowment from the Kavli Foundation and its founder Fred Kavli. IH is supported by a generous contribution from Philip Rice. DH and HX are supported by the Fermi Research Alliance, LLC under Contract No.~DE-AC02-07CH11359 with the U.S. Department of Energy, Office of Science, Office of High Energy Physics. ES is supported by the U.S. National Science Foundation Graduate Research Fellowship Program, grant number 2140001. GL is funded by the Grainger Graduate Fellowship from the University of Chicago Department of Physics.

\bibliography{LHAASO}

\clearpage

\onecolumngrid
\appendix

\section{Spectrum Power Law Index}

\begin{figure}[h!]
    \centering
    \begin{subfigure}{0.325\textwidth}
        \centering
        ~~~~~~~$\bm{\alpha = 2.7}$\par\medskip
        \includegraphics[width=\linewidth]{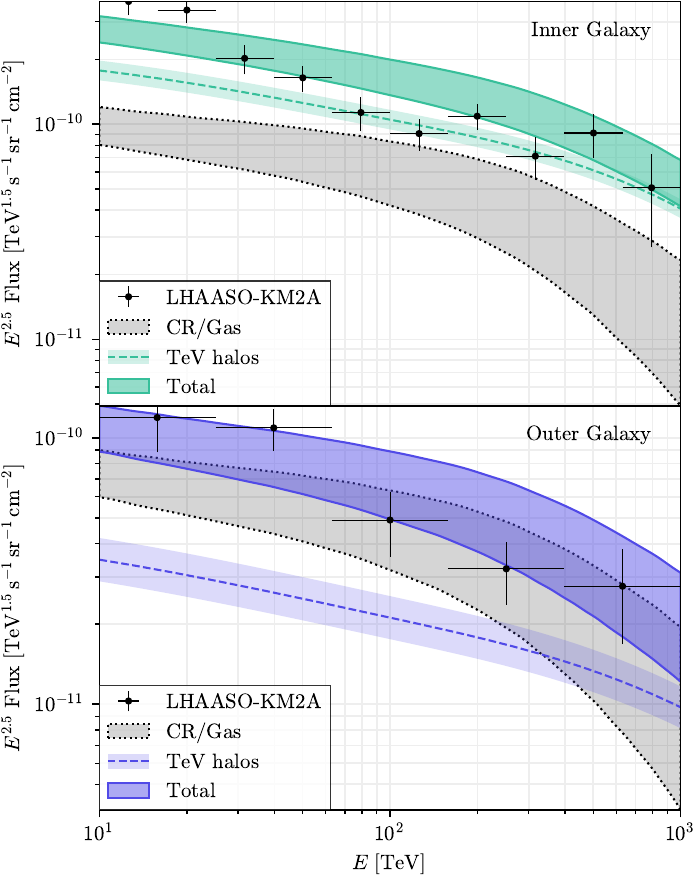}
    \end{subfigure}
    \begin{subfigure}{0.325\textwidth}
        \centering
        ~~~~~~~$\bm{\alpha = 3.1}$\par\medskip
        \includegraphics[width=\linewidth]{fig_spectrum.pdf}
    \end{subfigure}
    \begin{subfigure}{0.325\textwidth}
        \centering
        ~~~~~~~$\bm{\alpha = 3.4}$\par\medskip
        \includegraphics[width=\linewidth]{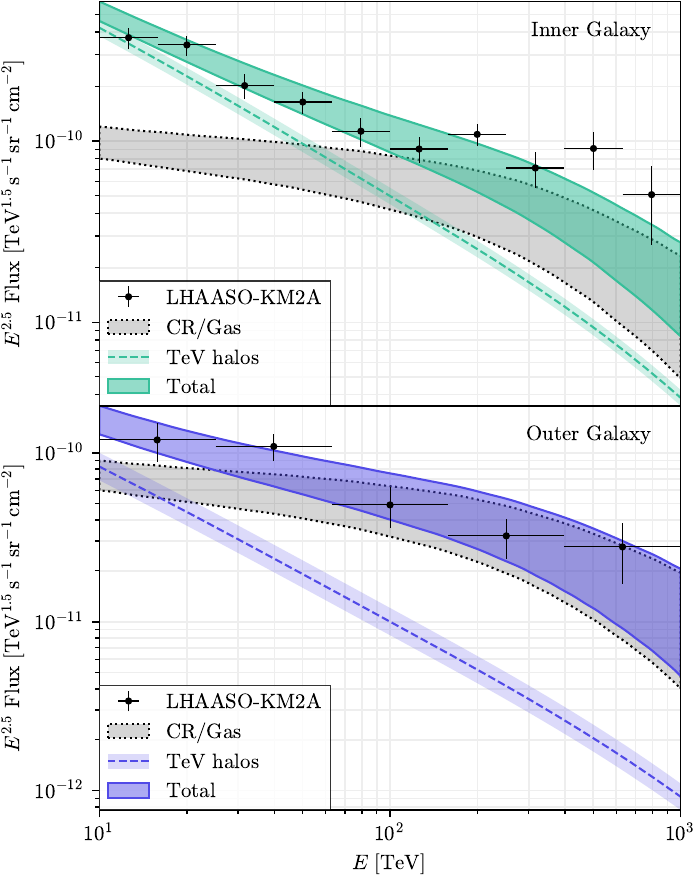}
    \end{subfigure}
    \caption{Same as Fig.~\ref{fig:spectrum} but for three different values of $\alpha$.}
    \label{fig:powerlawindex}
\end{figure}

Figure \ref{fig:powerlawindex} shows the spectral flux in the inner and outer galaxy from simulated TeV halos, cosmic-ray interactions with gas in the ISM, and their sum, for different values of the TeV halo power law index $\alpha$. These predictions are compared to the spectrum reported by LHAASO. Higher values of $\alpha$ fit the lower energy data points better, while lower $\alpha$ typically fits higher energy data points. Changing the spectral index also affects the TeV halo gamma ray efficiency $\eta$ required to account for the excess. For $\alpha = 2.7$, this requires $\eta = 0.02$, for $\alpha = 3.1$, $\eta = 0.052$, and for $\alpha = 3.4$, $\eta = 0.12$.

Throughout the paper, we use $\alpha = 3.1$ as a benchmark value that seems to provide a good fit to the LHAASO data. However, other data supports a range of possible $\alpha$ values for different TeV halo sources \cite{HAWC:2017kbo,HAWC:2020hrt,HAWC:2019tcx,HESS:2017lee}.

\end{document}